\documentclass[twocolumn,epsf,psfig]{revtex4-1}
\usepackage{amssymb}
\usepackage{epsfig}
\DeclareMathAlphabet{\pazocal}{OMS}{zplm}{m}{n}            
\usepackage{graphicx}
\usepackage{color}
\usepackage{bm}
\usepackage[normalem]{ulem}     
\usepackage{soul}
\usepackage{amsmath}
\usepackage{braket} 
\usepackage{rotating}
\usepackage{multirow} 
\DeclareMathAlphabet{\pazocal}{OMS}{zplm}{m}{n}            
\usepackage{graphicx}
\usepackage{hyperref}

\begin{document}
\title{Magnetoelectric classification of skyrmions} 

\author{Sayantika Bhowal} 
\affiliation{Materials Theory, ETH Zurich, Wolfgang-Pauli-Strasse 27, 8093 Zurich, Switzerland} 

\author{Nicola A. Spaldin}
\affiliation{Materials Theory, ETH Zurich, Wolfgang-Pauli-Strasse 27, 8093 Zurich, Switzerland}

\date{\today}

\begin{abstract}

We develop a general theory to classify magnetic skyrmions and related spin textures in terms of their magnetoelectric multipoles. Since magnetic skyrmions are now established in insulating materials, where the magnetoelectric multipoles govern the linear magnetoelectric response, our classification provides a recipe for manipulating the magnetic properties of skyrmions using applied electric fields. We apply our formalism to skyrmions and anti-skyrmions of different helicities, as well as to magnetic bimerons, which are topologically, but not geometrically, equivalent to skyrmions. We show that the non-zero components of the magnetoelectric multipole and magnetoelectric response tensors are uniquely determined by the topology, helicity and geometry of the spin texture. Therefore, we propose straightforward linear magnetoelectric response measurements as an alternative to Lorentz microscopy for characterizing insulating skyrmionic textures.

\end{abstract}

\maketitle

The concept of skyrmions, originally invoked by Skyrme to describe the stability of hadrons in particle physics more than half a century ago \cite{Skyrme1961,Skyrme1962}, has found fertile ground in condensed matter systems as diverse as liquid crystals \cite{Wright1989}, Bose-Einstein condensates \cite{Ho1998}, quantum Hall systems \cite{Sondhi1993}, and helimagnets \cite{Bogdanov1989,Rossler2006,Muhlbauer2009,Yu2010,Yu2011}. The magnetic skyrmions that form in the latter are metastable, topologically protected, nanometer-sized, swirling spin textures, with potential application as data bits in future high-density data storage devices \cite{Fert2013,Romming2013,Woo2016}.

\begin{figure}[t]
\centering
\includegraphics[width=\columnwidth]{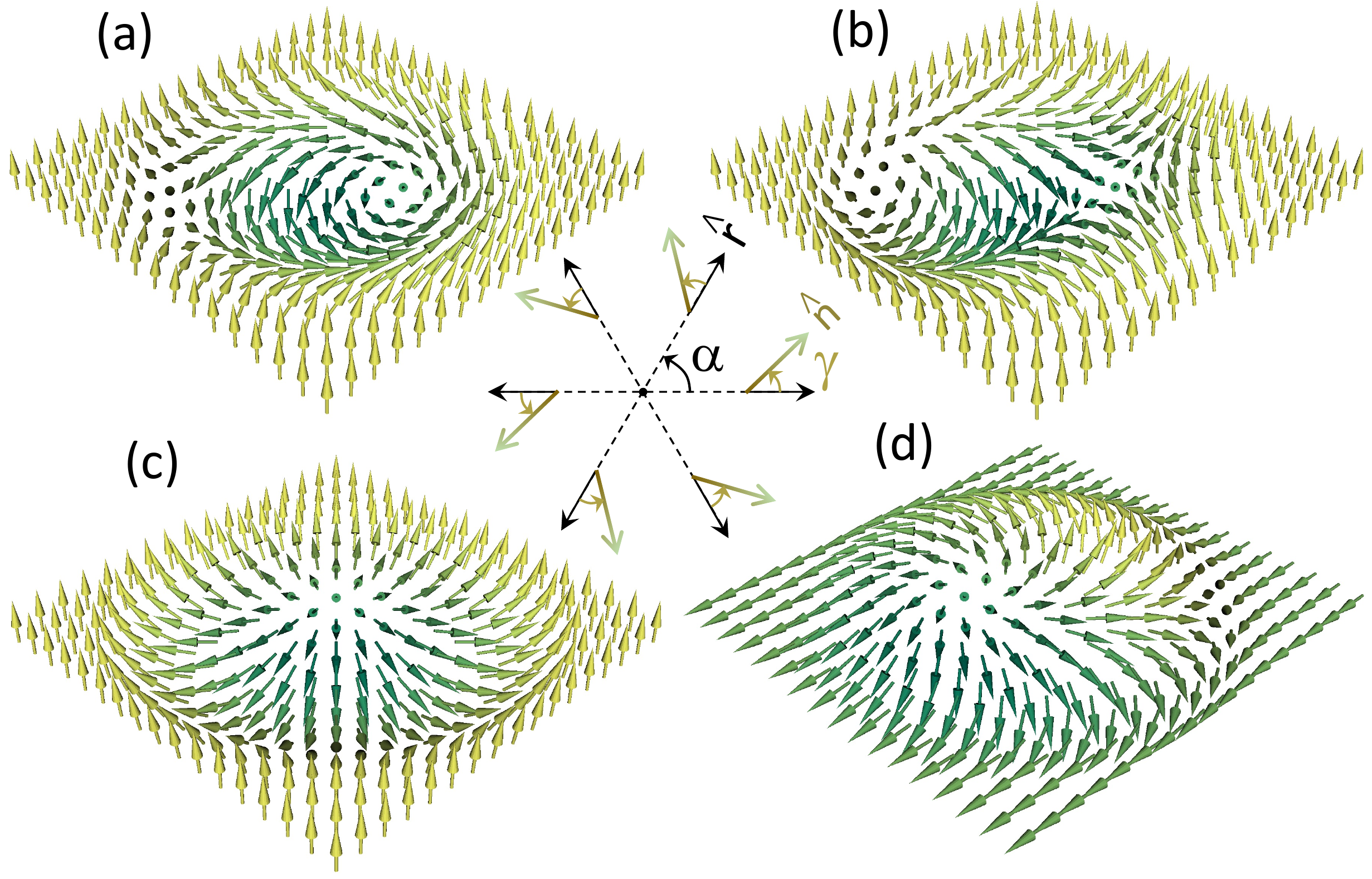}
 \caption{(a) Bloch skyrmion and (b) anti-skyrmion, (c) N\'eel skyrmion, and (d) bimeron. The schematic at the center illustrates the helicity $\gamma$, with gradient-colored arrows indicating the in-plane components of $\hat n$.
 }
 \label{fig-1}
 \end{figure}

Magnetic skyrmions have been found in bulk chiral and polar magnets \cite{Muhlbauer2009,Yu2010,Seki2012,Adams2012,Seki2012PRB,Kezsmarki2015,Fujima2017,Bordacs2017,Kurumaji2017}, thin film heterostructures \cite{Heinze2011,Woo2016,Romming2013}, and multilayer nanostructures,  \cite{Moreau-Luchaire2016,Wiesendanger2016,Leonov2016,Boulle2016,Soumyanarayanan2017,Zhang2018,Garlow2019}. Their radially symmetric spin texture is described by a local magnetization vector $\hat n (\theta(r),\phi(\alpha))$, with $\theta(r)$ and $\phi(\alpha)$ characterizing the radial profile and the twisting angle respectively, while the skyrmion number, $N_{sk} = \frac{1}{4\pi}\int dx dy ~\hat n \cdot (\partial_x \hat n \times \partial_y \hat n) $, characterizes the topology of the spin texture and manifests in the various exotic topological transport properties \cite{Neubauer2009,Nagaosa-Tokura2013,Matsuno2016}. For example, the topological invariants $N_{sk}=\pm 1$ represent the skyrmion and the anti-skyrmion respectively (see Figs \ref{fig-1} (a) and (b)).

In addition to their well-explored topological order, the lack of both space-inversion $\cal I$ and time-reversal $\tau$ symmetries in skyrmion-like spin textures makes them potential hosts for magnetoelectric (ME) multipoles \cite{ClaudeSpaldin2007,Spaldin2013}, formally defined as ${\cal M}_{ij} =  \int  r_i \mu_j (\vec r) d^3r$, with $\vec \mu (\vec r)$ being the magnetization density. The three irreducible (IR) components of the ${\cal M}_{ij}$ tensor, the ME monopole ($a$), toroidal moment ($\vec t$), and the ME quadrupole moment $q_{ij}$, are quintessential to the linear ME response $\alpha_{ij}$, which is the generation of magnetization (polarization) by an applied electric (magnetic) field. They have also been associated with other exciting properties and phases of matter, including hidden {\it ferrotoroidic} order \cite{Spaldin2008,Schmid,Aken}, 
current induced N\'eel vector switching in antiferromagnetic spintronics \cite{Watanabe2018,Florian2020}, and even with axionic dark matter \cite{Roising2021}. The recent observation \cite{Seki2012,Adams2012,Seki2012PRB,Kezsmarki2015,Fujima2017,Bordacs2017,Kurumaji2017} of skyrmions in insulators opens the door to combined electric and magnetic field manipulation of the skyrmions, mediated via these ME multipoles (MEMs).  
 \begin{table*} [t]
\caption{ME classification of skyrmions, anti-skyrmions and bimerons. Only independent multipolization components are listed.}
\setlength{\tabcolsep}{4pt}
\centering
\begin{tabular}{ c| c | c |   c| c| c|c}
\hline
Properties    \ &  \multicolumn{2}{c}{Skyrmions}  \ & \multicolumn{2}{c}{Anti-skyrmions} \ &  \multicolumn{2}{c}{Bimerons}    \\
              \ &     $\gamma=0$ (N\'eel) \ &   $\gamma=\pi/2$ (Bloch) & $\gamma=0$  \ &   $\gamma=\pi/2$ & $\gamma=0$  \ &   $\gamma=\pi/2$\\ [1 ex]
\hline
ME multipolization & ${\cal A}$ $\propto \cos \gamma$ & ${\cal T}_z$ $\propto \sin \gamma$ & ${\cal Q}_{x^2-y^2}$ $\propto \cos \gamma$ & ${\cal Q}_{xy}$ $\propto \sin \gamma$ & ${\cal Q}_{x^2-y^2}$ $\propto \cos \gamma$ & ${\cal T}_x$ $\propto \sin \gamma$\\
ME polarizability & $\alpha_{xx} = \alpha_{yy}$ & $\alpha_{xy} = -\alpha_{yx}$ & $\alpha_{xx} =-\alpha_{yy}$ & $\alpha_{xy} = \alpha_{yx}$ & $\alpha_{xz} = -\alpha_{yy}$& $\alpha_{xy} = \alpha_{yz}$\\
  \hline
\end{tabular}
\label{tab1}
\end{table*}

In spite of this intriguing connection, to our knowledge, the only link between MEMs and skyrmions mentioned to date, is to their toroidization ${\cal T}_z$ (toroidal moment per unit volume, $\vec t/V$) \cite{Gobel2019}. Here we present a complete ME classification of skyrmions and anti-skyrmions with different helicities, $\gamma$, (see Fig. \ref{fig-1}) as well as magnetic bimerons [see Fig. \ref{fig-1} (d)] taking all components of the magnetoelectric multipolization (that is the MEM per unit volume) into account. We find that the ME monopolization ${\cal A}$ (ME monopole per unit volume, $a/V$) and the quadrupolization ${\cal Q}_{ij}$ (quadrupole moment per unit volume, $q_{ij}/V$) can be non-zero, in addition to the toroidization $\vec {\cal T}$, with the 
form of the MEM tensor depending on both the topology and geometry of the spin texture. These distinct MEMs in skyrmions, anti-skyrmions, and bimerons, further, manifest in the corresponding ME polarizability $\alpha_{ij}$, implying that the skyrmion type can be determined from a straightforward magnetoelectric measurement, and pointing to combined electric- and magnetic-field control of skyrmions. 

The main outcomes of the classification are given in Table \ref{tab1}, which summarizes the key findings of our work. First, while skyrmions with different helicities carry the same topological order ($N_{sk}$) and are, therefore, topologically indistinguishable, they have different MEMs, which, in turn, leads to different ME polarizability $\alpha_{ij}$. As a result, the ME polarizability can be used as an alternative to Lorentz microscopy \cite{Yu2010}
to classify skyrmions with different helicities. Second, skyrmions and anti-skyrmions differ not only in their topological order, but also have different MEMs, and, hence, different ME response. Therefore skyrmions and antiskyrmions can be characterized and detected in experiments using a single observable $\alpha_{ij}$. The case of the bimeron emphasizes the strong dependence of the form of the MEM and $\alpha_{ij}$ tensors on the geometry of the spin texture. Finally, the unique spin textures of skyrmions and anti-skyrmions facilitate the existence of multipolization components in their purest form. For example, Bloch skyrmions have a pure toroidization ${\cal T}_z$, anti-skyrmions with $\gamma=\pi/2$ have a pure quadrupolization ${\cal Q}_{xy}$, etc. Such a pure toroidal or quadrupolar state is rather rare in materials with magnetic order on the unit-cell scale, where they often coexist with each other due to symmetry \cite{Spaldin2008,Spaldin2013,Spaldin2021,BhowalSpaldin2021,Bhowal2021}.      

{\it ME multipolization and spin geometry in two dimensions.} 
Here we briefly review the ME multipolization and discuss the dependence of the multipolization tensor on the lattice dimension and the spin geometry of the usual {\it two dimensional} (2D) topological skyrmion-like spin texture.

The ME multipolization ${\tilde {\cal M}}_{ij} = {\cal M}_{ij}/V$, i.e., the MEM moment per unit volume $V$, describes the first order asymmetry in the magnetization density $\vec \mu (\vec r)$ that couples to derivatives of the magnetic field \cite{Spaldin2013,Gao2018}. Hereafter, we only consider the spin part of the magnetic moment. As stated earlier, the ${\cal M}_{ij}$ tensor has three IR components \cite{Spaldin2013}: (a) the scalar ME monopole $a= \frac{1}{3} {\cal M}_{ii} =  \int  \vec r \cdot \vec \mu (\vec r) d^3r$, (b) the ME toroidal moment  $t_i= \frac{1}{2} \varepsilon_{ijk} {\cal M}_{jk} =  \frac{1}{2} \int  \vec r \times \vec \mu (\vec r) d^3r$, and (c) the symmetric traceless five component quadrupole moment tensor $q_{ij}=\frac{1}{2} ( {\cal M}_{ij}+ {\cal M}_{ji} -\frac{2}{3} \delta_{ij} {\cal M}_{kk})= \frac{1}{2}   \int  \big(r_i \mu_j + r_j \mu_i - \frac{2}{3} \delta_{ij} \vec r \cdot \vec \mu \big)  d^3r$. The ME monopole and the $q_{x^2-y^2}$ and $q_{z^2}$ quadrupole moment components form the diagonal of the ${\cal M}_{ij}$ tensor, while the symmetric and the anti-symmetric parts of the off-diagonal elements are represented by the $q_{xy},q_{xz},$ and $q_{yz}$ quadrupole moments and $\vec t$ respectively. 

\begin{figure}[t]
\centering
\includegraphics[width=\columnwidth]{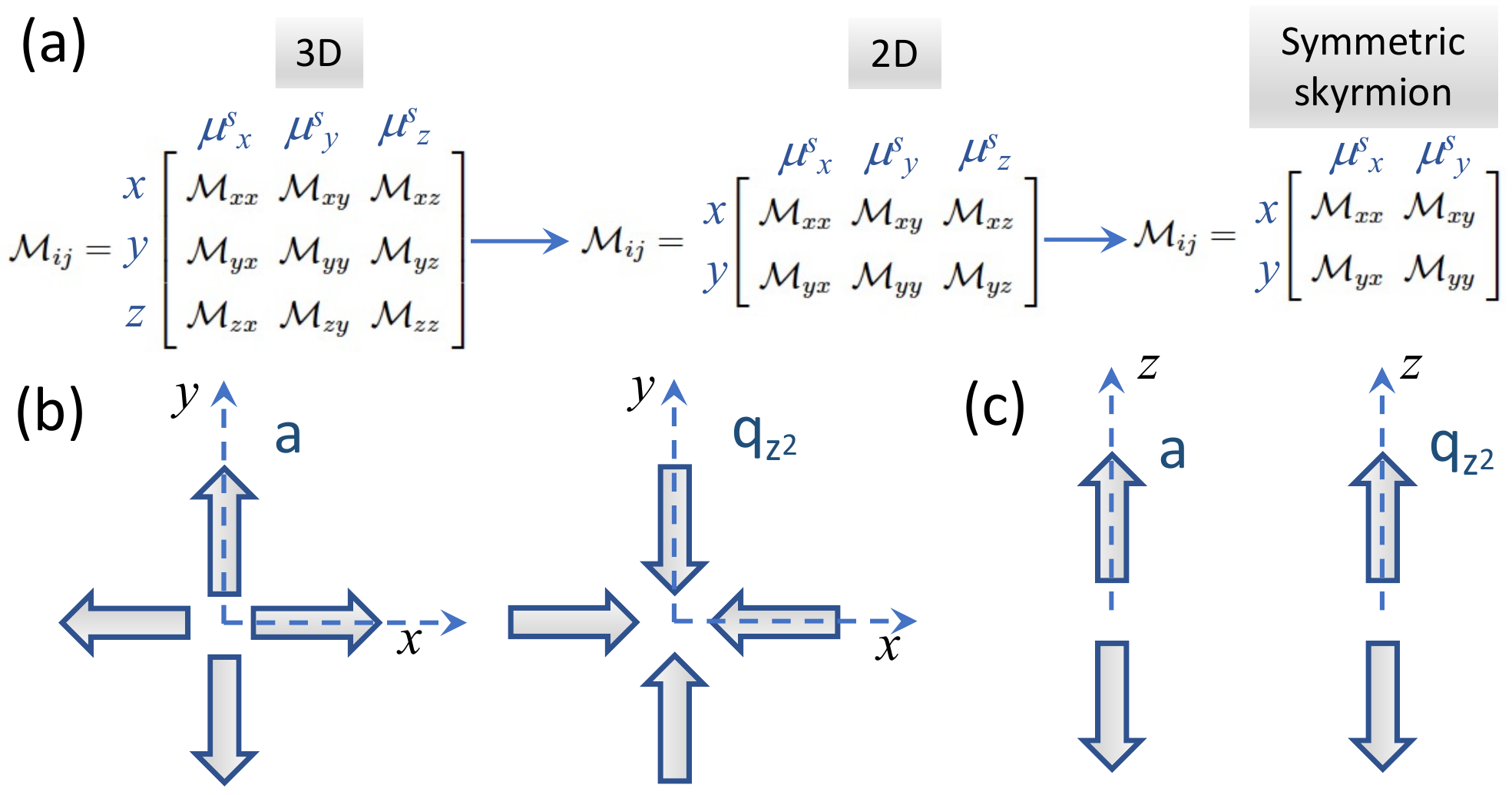}
 \caption{(a) Schematics showing the evolution of the ${\cal M}_{ij}$ tensor as the structural dimension reduces from $3D$ to $2D$, and finally for the specific radially symmetric skyrmion spin texture. Representative spin magnetic moment arrangements (shown in arrows) for a ME monopole $a$ and the quadrupole moment $q_{z^2}$ in the (b) $x$-$y$ plane and (c) along $z$, showing that $a$ and $q_{z^2}$ in (b) are exactly equal and opposite in the absence of a local $z$ coordinate.}
 \label{fig0}
 \end{figure}

We now point out some interesting features of the ${\cal M}_{ij}$ tensor based on  the reduced lattice dimension from $3D \rightarrow 2D$, and the spin geometry. 
First, a $2D$ in-plane lattice directly implies vanishing ${\cal M}_{zi} = \int z\mu_i d^2r$ components, and the corresponding multipolization (which in this case is multipole moment per unit area $S$, ${\tilde {\cal M}}_{ij} = {\cal M}_{ij}/S$). Here $i=x,y,$ and $z$ are the Cartesian components of $\vec \mu$. Moreover, the radially symmetric $\mu_z$ spin component at each lattice site of a skyrmion crystal forces the ${\cal M}_{iz} = \int r_i\mu_z d^2r$ components to vanish. 
The resulting ${\cal M}_{ij}$ tensor can therefore be written as a $2\times2$ matrix of non-zero components [see Fig. \ref{fig0} (a)]. In contrast, the absence of $\mu_z$ radial symmetry in a bimeron texture means that the ${\cal M}_{ij}$ tensor does not reduce to a $2\times2$ matrix. The different dimensionality of their ${\cal M}_{ij}$ tensors emphasizes that although topologically equivalent, skyrmions and bimerons have distinct ME responses originating from geometrical differences. Secondly, in a $2D$ lattice, the ME monopole $a$ and the quadrupole $q_{z^2}$ are equal and opposite to each other: $a=\frac{1}{3} \int d^2r (x\mu_x+y\mu_y+z\mu_z)=\frac{1}{3} \int d^2r (x\mu_x+y\mu_y)$, and $q_{z^2}=\frac{1}{2} \int d^2r  \{ z\mu_z+z\mu_z-\frac{2}{3}(x\mu_x+y\mu_y) \} = -\frac{1}{3} \int d^2r (x\mu_x+y\mu_y)=-a$. This means that in a $2D$ lattice 
$q_{z^2}$ (${\cal Q}_{z^2}$) is the same as an anti-$a$ (anti-${\cal A}$) and vice versa (see Figs. \ref{fig0} (b) and (c)).

{\it Results and discussion.}
We begin by considering a tight-binding model for an electron in a square lattice of spin texture $\hat n_i$ \cite{Hamamoto2015},
\begin{equation} \label{tb}
{\cal H} = t \sum_{\langle i, j \rangle} c_i^\dagger c_j -J_{\rm H} \sum_i \hat n_i \cdot (c_i^\dagger \vec \sigma c_i).
\end{equation}
Here $t$ and $J_{\rm H}$ are the nearest-neighbor hopping and Hund's coupling respectively. For a skyrmion texture, $\hat n_i \equiv ( \sin\theta_i(r_i)\cos\phi_i(\alpha_i),  \sin\theta_i(r_i)  \sin\phi_i(\alpha_i),\cos\theta_i(r_i))$ with $\theta_i=\pi (1-r_i/\lambda)$ \cite{Nagaosa-Tokura2013} and $\phi_i = m \alpha_i +\gamma$. Here, the vorticity $m =\pm 1$ corresponds to a skyrmionic and an anti-skyrmionic state respectively while the helicity $\gamma$ can take different values; e.g., $\gamma=0$ and $\pi/2$ correspond to Bloch and  N\'eel  skyrmions respectively [see Figs \ref{fig-1} (a) and (c)]. The bimeron configuration, also known as in-plane magnetized version of a skyrmion [see Fig. \ref{fig-1} (d)], is obtained from the skyrmion texture via spin rotation by $\pi/2$ around the $y$ axis, $(\hat n_x, \hat n_y, \hat n_z) \rightarrow (\hat n_z, \hat n_y, -\hat n_x)$, losing thereby the radial symmetry of the spin-$z$ component of a skyrmion while keeping the topology intact \cite{Kharkov2017,GobelPRB2019}.  

In the adiabatic limit ($J_{\rm H} \gg t$), the low lying bands of the tight-binding model, Eq. (\ref{tb}), can be approximated as \cite{Hamamoto2015}
\begin{equation} \label{tbeff}
{\cal H} = \sum_{\langle i, j \rangle}  t_{ij}^{\rm eff}d_i^\dagger d_j ,
\end{equation}
where $d_i, d_i^\dagger$ are spin-less operators, and $t_{ij}^{\rm eff}=t \cos (\tilde{\theta}_{ij}/2) e^{ia_{ij}}$ is an effective hopping that depends on the twisting angle difference $(\phi_i-\phi_j)$ \cite{Skphase}. 

\begin{figure}[t]
\centering
\includegraphics[width=\columnwidth]{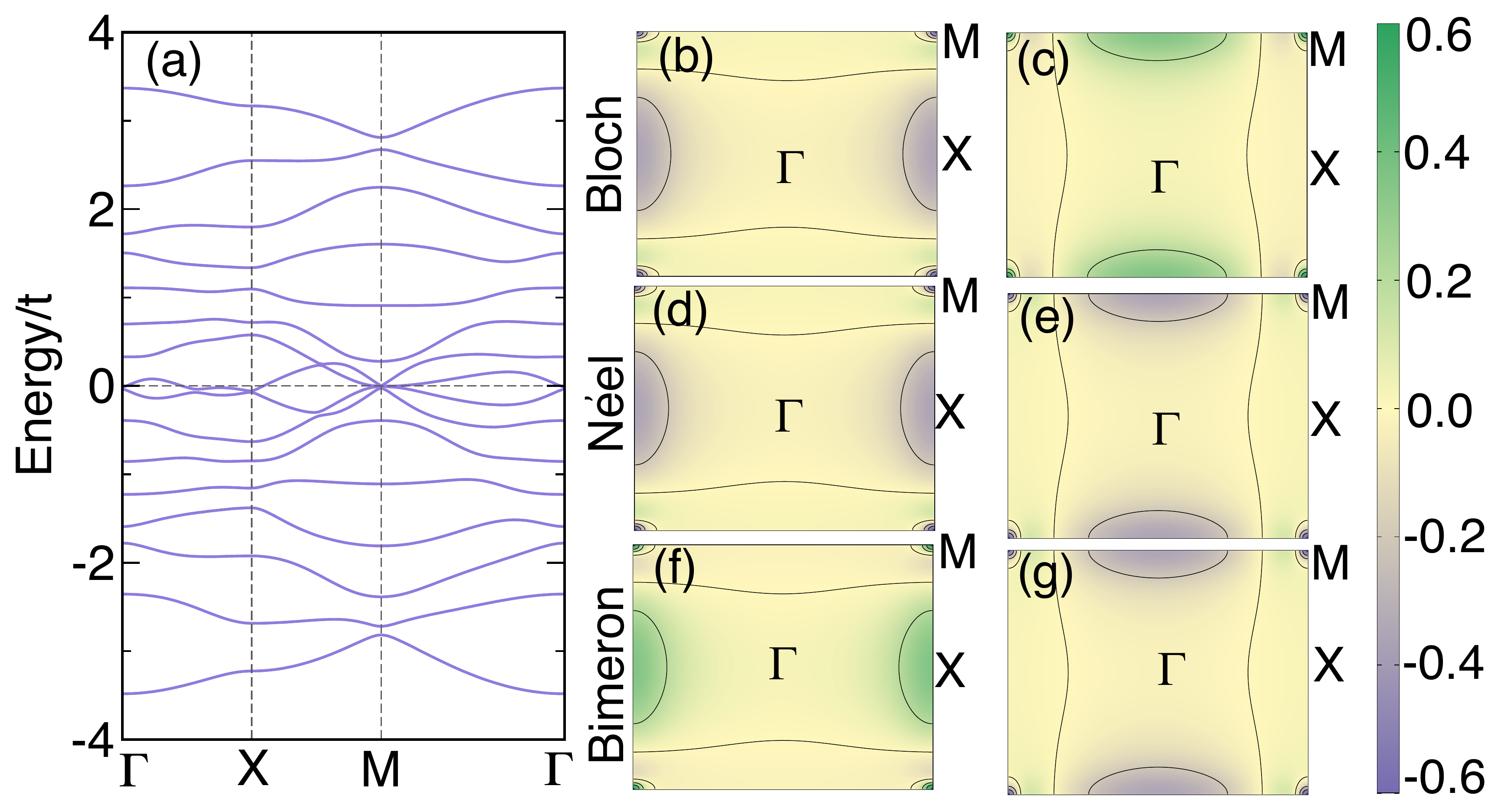}
 \caption{(a) Band structure for the model Hamiltonian, Eq. (\ref{tb}) with parameter $J_{\rm H}/t =10$. For simplicity, the lowest sixteen bands of the model Hamiltonian (\ref{tb}) around the energy $E=-J_{\rm H}/t$ are shown and the energies are shifted by $E$, indicated by the horizontal dashed line. (b)-(g) Momentum space distribution of ${\cal O}_{ij} (\vec k)$ of Eq. (\ref{MEM}). (b) ${\cal O}_{xy}(\vec k)$ and (c) ${\cal O}_{yx}(\vec k)$ for the Bloch skyrmion; (d) ${\cal O}_{xx}(\vec k)$ and (e) ${\cal O}_{yy}(\vec k)$ for the N\'eel skyrmion; (f) ${\cal O}_{xz}(\vec k)$ and (g) ${\cal O}_{yy}(\vec k)$ for the bimeron. The Fermi energy is taken to be at the top of the lowest band in (a).}
 \label{fig2}
 \end{figure}

We first analyze the band structures of $2D$ periodic crystals of skyrmions, bimerons, and antiskyrmions, computed from the tight-binding model, Eq. (\ref{tb}). The low-lying bands for $J_{\rm H}/t = 10$, shown in Fig \ref{fig2}(a), are well described by the adiabatic limit Hamiltonian of Eq. (\ref{tbeff}) and are identical for the three topological spin textures for a given set of parameters. This is because the effective hopping $t_{ij}^{\rm eff}$ depends only on the difference in $\phi$, and, therefore, remains the same for any constant rotation of the spins. Since N\'eel and Bloch skyrmions differ only in helicity $\gamma$, and bimerons by a $\frac{\pi}{2}$ rotation around $y$-axis, they therefore have the same $t_{ij}^{\rm eff}$. For anti-skyrmions, $t_{ij}^{\rm eff}$ has the opposite phase leading to opposite topological order while keeping the band energies unaltered.

 While the bandstructure is insensitive to rotations of the spins, they manifest in the  corresponding spin multipolization $\tilde{{\cal M}}_{ij}$, which can be computed as the Brillouin zone (BZ) integration over all the occupied states $n$ of ${\cal O}^n_{ij} (\vec k)$, \cite{Gao2018,symnote}
\begin{eqnarray} \label{MEM} \nonumber
\tilde{{\cal M}}_{ij} &=& -g\mu_B \int^{occ} \frac{d^2k}{(2\pi)^2} \sum_n {\cal O}^n_{ij} (\vec k),~  {\rm where}~\\ 
{\cal O}^n_{ij}(\vec k) &=&  \sum_{m \ne n} (\varepsilon_n+\varepsilon_m-2\varepsilon_F) {\rm Im} \Big[\frac{\langle n|v_i| m\rangle \langle m |s_j| n \rangle}{ (\varepsilon_n-\varepsilon_m)^2} \Big].
\end{eqnarray}
Here $v_i$ and $s_i$ are the velocity and the Pauli spin operators respectively.

We compute the $\tilde{{\cal M}}_{ij}$ tensor for $\gamma$ values ranging from $-\pi$ to $\pi$, assuming that only the lowest band in Fig. \ref{fig2} (a) is occupied. 
 We start with the case of the skyrmion crystal, and discuss pure Bloch- and N\'eel- type skyrmions first, before analyzing intermediate $\gamma$ values. 
Our calculations show that the Bloch skyrmions have only non-zero off-diagonal elements, $\tilde{{\cal M}}_{xy} = -\tilde{{\cal M}}_{yx}$, indicating the presence of only the toroidization ${\cal T}_z$, consistent with Ref. \cite{Gobel2019}. 
We note that a pure ${\cal T}_z$ is unusual 
\cite{Spaldin2008,Spaldin2013,Spaldin2021,BhowalSpaldin2021,Bhowal2021}; while recently we demonstrated a pure toroidal moment in the reciprocal space of PbTiO$_3$ \cite{BhowalCollinsSpaldin}, to the best of our knowledge this is the first prediction of a pure toroidal moment in a real-space spin texture. In complete contrast, N\'eel skyrmions have only non-zero diagonal elements, $\tilde{{\cal M}}_{xx} = \tilde{{\cal M}}_{yy}$.    
Since the ME monopolization $\cal A$, and quadrupolization ${\cal Q}_{x^2-y^2}$ and ${\cal Q}_{z^2}$ contribute to the diagonal elements as
\begin{eqnarray} \nonumber \label{diag}
\tilde{{\cal M}}_{xx} &=& {\cal A} +\frac{1}{2}({\cal Q}_{x^2-y^2}- {\cal Q}_{z^2}) \\ \nonumber
\tilde{{\cal M}}_{yy} &=& {\cal A} -\frac{1}{2}{\cal Q}_{x^2-y^2}- \frac{1}{2}{\cal 
Q}_{z^2} \\
\tilde{{\cal M}}_{zz} &=& {\cal A}+ {\cal Q}_{z^2},
\end{eqnarray}
 this implies the presence of a ME monopolization ${\cal A}=-{\cal Q}_{z^2}$ in a N\'eel skyrmion, consistent with our previous discussion that $a = -q_{z^2}$ in a $2D$ system. 

In Figs. \ref{fig2} (b)-(e), we show the $k$-space distributions of ${\cal O}_{ij} (\vec k)$ for Bloch and N\'eel skyrmions.
It is interesting to point out the multipolization symmetries, ${\cal O}^{\rm Bloch}_{xy} (\vec k) = {\cal O}^{\text{\rm N\'eel}}_{xx} (\vec k)$, ${\cal O}^{\text{\rm Bloch}}_{yx} (\vec k) = -{\cal O}^{\text{\rm N\'eel}}_{yy} (\vec k)$, which follow from the differences in $\gamma$: $\mu_y^{\rm Bloch}= \sin \theta \sin (\pi/2 + \phi)=\mu_x^{\text{\rm N\'eel}}$. Furthermore, since the velocity $v_x$ does not depend on $\gamma$, it is easy to see from Eq. (\ref{MEM}) that ${\cal O}^{\rm Bloch}_{xy} = {\cal O}^{\text{\rm N\'eel}}_{xx}$. Similarly, it can be shown that $\mu_x^{\rm Bloch} =-\mu_y^{\text{\rm N\'eel}}$, leading to ${\cal O}^{\rm Bloch}_{yx} (\vec k) = -{\cal O}^{\text{\rm N\'eel}}_{yy} (\vec k)$. Note that, although the explicit value of ${\cal O}_{ij}$ depends on the Fermi energy $\varepsilon_F$ [see Eq. (\ref{MEM})], the relations discussed here remain intact because they are determined by the symmetries of the spin texture.

\begin{figure}[t]
\centering
\includegraphics[width=\columnwidth]{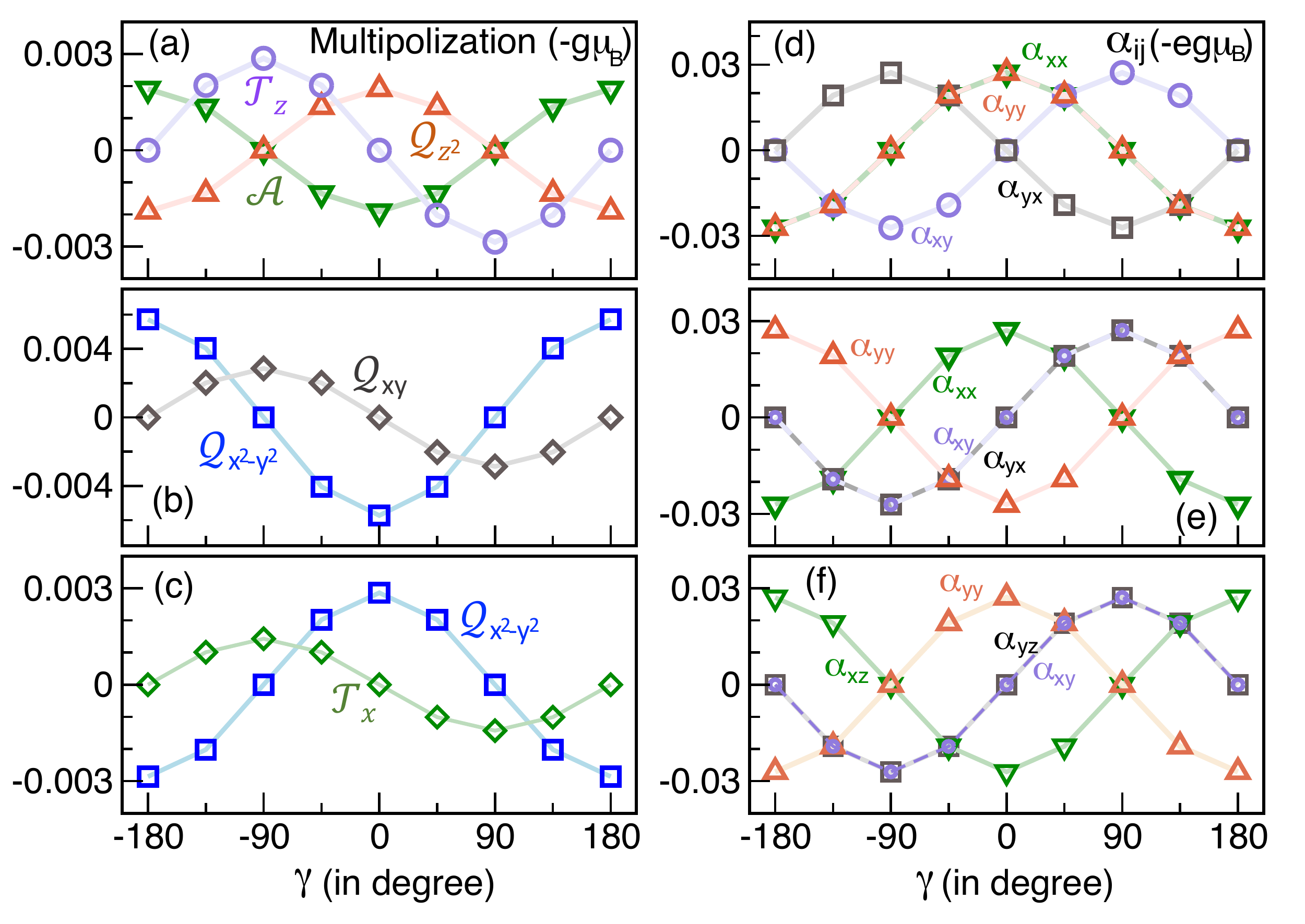}
 \caption{Variation in the ME multipolization components as a function of the helicity $\gamma$ in (a) skyrmions, (b) anti-skyrmions, and (c) bimerons. The same variation for the ME polarizability components $\alpha_{ij}$ in (d) skyrmions, (e) anti-skyrmions, and (f) bimerons.}
 \label{fig3}
 \end{figure}

Finally, for twisted skyrmions described by intermediate $\gamma$ values, 
we find that both diagonal and off-diagonal elements of ${\tilde {\cal M}}_{ij}$ exist, resulting in non-zero ${\cal T}_z$ and ${\cal A}=-{\cal Q}_{z^2}$, that vary periodically with $\gamma$ [see Fig. \ref{fig3} (a)]. While ${\cal T}_z$ varies as $\sin \gamma$, ${\cal A}$ (${\cal Q}_{z^2}$) varies as $\cos \gamma$. These periodic dependences of ${\cal T}_z$ and ${\cal A}$ (${\cal Q}_{z^2}$) can be understood from their formal definitions. For example, ${\cal T}_z= t_z/S,$ with $t_z=\int d^2r (\vec r \times \vec \mu)$. For skyrmions $\vec \mu \equiv (\sin \theta \cos \phi, \sin \theta \sin \phi, \cos \theta)$, with $\phi(\alpha)=\alpha + \gamma$, and $(x,y)\equiv r(\cos \alpha, \sin \alpha)$. Substituting this expression for $\vec \mu$, we obtain ${\cal T}_z \propto \sin \gamma$. Similarly, the ME monopolization ${\cal A}=a/S=\frac{1}{S} \int d^2r (\vec r \cdot \vec \mu) \propto \cos \gamma$. 

The spin multipolization, discussed above, is directly related to the spin ME polarizability, $\alpha_{ij}=-e\frac{\partial \tilde{{\cal M}}_{ij}}{\partial \varepsilon_F}$, which can, therefore, be computed as \cite{Gao2018}, 
\begin{eqnarray} \label{alpha} \nonumber
\alpha_{ij} &=& e g\mu_B \int^{occ} \frac{d^2k}{(2\pi)^2} \sum_n {\cal D}^n_{ij} (\vec k),  \\ 
{\rm where}~~ {\cal D}^n_{ij}(\vec k) &=&  -2~ {\rm Im} \sum_{m \ne n}  \Big[\frac{\langle n|v_i | m\rangle \langle m |s_j| n \rangle}{ (\varepsilon_n-\varepsilon_m)^2} \Big].
\end{eqnarray}
Here the integration is over the occupied part of the BZ. We compute the response $\alpha_{ij}$ for the skyrmion as a function of $\gamma$ and show our results in Fig. \ref{fig3} (d). Similar to the $\tilde{{\cal M}}_{ij}$ tensor, we find that $\alpha_{ij}$ has a $2\times2$ matrix form, with $\alpha_{xx}=\alpha_{yy}$ varying as $\cos \gamma$, and $\alpha_{xy}=-\alpha_{yx}$ as $\sin \gamma$ [see Fig. \ref{fig3} (d)], reflecting the one-to-one correspondence between $\tilde{{\cal M}}_{ij}$ and $\alpha_{ij}$. A similar sine/cosine dependence of $\alpha_{ij}$ was predicted earlier for magnetic vortices \cite{Delaney2009}.  

We now turn to the case of an antiskyrmion with vorticity $m=-1$. The variations of $\tilde{{\cal M}}_{ij}$ and $\alpha_{ij}$ with $\gamma$ in an antiskyrmion are shown in Figs. \ref{fig3} (b) and (e). Similarly to the case of skyrmions, at $\gamma=0$, the $\tilde{{\cal M}}_{ij}$ tensor of an antiskyrmion has a diagonal $2\times2$ matrix form, but the diagonal elements have opposite signs, $\tilde{{\cal M}}_{xx} =-\tilde{{\cal M}}_{yy}$. Following Eq. (\ref{diag}), this implies the presence of quadrupolization ${\cal Q}_{x^2-y^2}$, which, as seen from Fig. \ref{fig3} (b), varies as $\cos \gamma$, with its maximum at $\gamma=0$ and zero at $\gamma=\pi/2$. At $\gamma=\pi/2$, the $\tilde{{\cal M}}_{ij}$ tensor is off-diagonal with $\tilde{{\cal M}}_{xy}=\tilde{{\cal M}}_{yx}$, in contrast to the opposite signs in skyrmions, leading to a quadrupolization ${\cal Q}_{xy}$ that varies as $\sin \gamma$. These periodic dependences again follow from the formal definitions: ${\cal Q}_{x^2-y^2}=\frac{1}{S}\int d^2r (x\mu_x-y\mu_y) \propto \cos \gamma$, noting that $\phi=-\alpha+\gamma$ for an anti-skyrmion, and ${\cal Q}_{xy}=\frac{1}{2S} \int d^2r (x\mu_x+y\mu_y) \propto \sin \gamma$. The corresponding computed polarizability $\alpha_{ij}$ [Fig. \ref{fig3} (e)] follows the $\tilde{{\cal M}}_{ij}$ tensor, with $\alpha_{xx}=-\alpha_{yy} \propto \cos \gamma$ and $\alpha_{xy}= \alpha_{yx} \propto \sin \gamma$.      

Finally, we discuss the bimeron crystal, motivated by its closely related spin texture to that of skyrmions. Despite being topologically equivalent to skyrmions, the lack of radial symmetry in the $\mu_z$ components of a bimeron texture results in a very different form of the $\tilde{{\cal M}}_{ij}$ tensor, emphasizing the crucial dependence on spin geometry. In contrast to the skyrmions and anti-skyrmions described above, the bimeron $\tilde{{\cal M}}_{ij}$ tensor has a $2\times3$ matrix form that corresponds to the middle panel of Fig. \ref{fig0} (a), with non-zero elements    $\tilde{{\cal M}}_{xz}=-\tilde{{\cal M}}_{yy}$ at $\gamma=0$. The non-zero elements of the $\tilde{{\cal M}}_{ij}$ tensor can be understood by noting that $\mu_z^{\rm bimeron} =-\mu_x^{\text{\rm N\'eel}}$, while $\mu_y^{\rm bimeron} =\mu_y^{\text{\rm N\'eel}}$. Since the velocities $v_x$ and $v_y$ are the same for both textures (they have identical band structures, Fig. \ref{fig2} (a)), this implies $\tilde{{\cal M}}^{\rm bimeron}_{xz}=-\tilde{{\cal M}}^\text{\rm N\'eel}_{xx}$ and $\tilde{{\cal M}}^{\rm bimeron}_{yy}=\tilde{{\cal M}}^\text{\rm N\'eel}_{yy}$, which is also reflected in the corresponding $k$-space distributions of ${\cal O}_{ij} (\vec k)$, shown in Fig. \ref{fig2} (f) and (g) respectively. Consequently, $\tilde{{\cal M}}_{xx}=\tilde{{\cal M}}_{yy}$ in a N\'eel skyrmion translates into $\tilde{{\cal M}}_{xz}=-\tilde{{\cal M}}_{yy}$ in a bimeron. This further implies the presence of 
${\cal Q}_{x^2-y^2}=3{\cal Q}_{z^2}=-3{\cal A}=-2{\cal T}_{y}=2{\cal Q}_{xz}$ at $\gamma=0$. All the MEMs vary as $\cos \gamma$ with zero value at $\gamma=\pi/2$, at which non-zero elements are $\tilde{{\cal M}}_{xy}=\tilde{{\cal M}}_{yz}$, indicating the presence of ${\cal T}_{x}={\cal Q}_{yz}={\cal T}_{z}={\cal Q}_{xy} \propto \sin \gamma$. The variation of the independent multipolization components with $\gamma$ is shown in Fig. \ref{fig3} (c). A similar periodic dependence is also evident in the corresponding polarizability $\alpha_{ij}$, shown in Fig. \ref{fig3} (f). 

To summarize, we have introduced a ME classification of skyrmions and applied it to Bloch and N\'eel skyrmions \cite{TokuraNaoya2021}, as well as anti-skyrmions, and bimerons. The formalism is general and can be extended to other spin textures. Our work opens the door for future works examining the implications of MEMs in both metallic and insulating skyrmion-like textures. In particular, dependence of the multipolization on the helicity demonstrated here, may have useful implications in the context of recent efforts to control the helicity of skyrmions \cite{Bo2021,Shibata2013}. 

Note that the ME monopole associated with a N\'eel skyrmion in the present work is different from the previously predicted magnetic monopole at the point of coalescence of two in-going skyrmion lines \cite{Milde2013}. In particular, unlike the ME monopole, the magnetic monopole does not break inversion symmetry, and appears in the zeroth order term of the multipole expansion of a magnetization density in a magnetic field. However, care should be taken as often in the literature this distinction based on formal definition is not followed. For example, as pointed out by Khomskii \cite{Khomskii2012,Khomskii2014}, the elementary excitation in spin ice that is referred to as a magnetic monopole  carries an electric dipole (and so breaks space-inversion symmetry) in addition to a magnetic charge.

We hope that our work stimulates experimental efforts in ME manipulation of skyrmions and related topological spin textures, opening up new avenues in designing unique {\it skyrmionic} (skyrmion-based spintronic) devices with high energy efficiency.

\section*{Acknowledgements}
NAS and SB were supported by the ERC under the EU’s Horizon 2020 Research and Innovation Programme grant No 810451 and by the ETH Zurich. Computational resources were provided by ETH Zurich's Euler cluster, and the Swiss National Supercomputing Centre, project ID eth3.

\bibliographystyle{apsrev4-1}
\bibliography{LNPO}

\begin{thebibliography}{59}
\bibitem [{sym()}]{symnote}%
  \BibitemOpen
  \href@noop {} {}\bibinfo {note} {Note that under both time reversal and space
  inversion, the integrand ${\cal O}^n_{ij}(\vec k) {\rightarrow} -{\cal
  O}^n_{ij}(-\vec k)$, implying that a non-zero value of the integral in Eq.
  (\ref{MEM}) requires both symmetries to be broken.}\BibitemShut {Stop}%
\end{thebibliography}

\end{document}